\documentclass[prd,showpacs,onecolumn,preprintnumbers,preprint ]{revtex4}
\usepackage{amsmath} \usepackage{graphicx} \usepackage{amsfonts}
\usepackage{array} \usepackage{amsthm}

\usepackage[dvips]{epsfig}
\usepackage{psfrag}

\newcommand{\vf}{\varphi} \newcommand{\nn}{\nonumber}
\newcommand{\qd}{\qquad} 
 
\newcommand{\Lef}{\left(} \newcommand{\Rig}{\right)}
\newcommand{\p}{\partial} 
 
 \newcommand{\8}{S_\infty}
\newcommand{\1}{\Psi_1}

\newcommand{\0}{p_0}

\newcommand{\ta}{\tilde{\alpha}}
\newcommand{\ra}{\rightarrow}
\begin{document}
\preprint{DTP-MSU/08-23}
\title{Curvature-corrected dilatonic black holes and
black hole -- string transition}
\author{
  Dmitri V. Gal'tsov}
\email{galtsov@phys.msu.ru}\affiliation{Department of Theoretical
Physics, Moscow State University, 119899, Moscow, Russia}
\author{
    Evgeny A. Davydov}
\email{eugene00@mail.ru} \affiliation{Department of Theoretical
Physics, Moscow State University, 119899, Moscow, Russia}
\date{00.00.00}
\begin{abstract}
We investigate extremal charged black hole solutions in the
four-dimensional string frame Gauss-Bonnet gravity with the Maxwell
field and the dilaton. Without curvature corrections, the extremal
electrically charged dilatonic black holes have singular horizon and
zero Bekenstein entropy. When the Gauss-Bonnet term is switched on,
the horizon radius expands to a finite value provided  curvature
corrections are strong enough. Below a certain threshold value of
the Gauss-Bonnet coupling the extremal black hole solutions cease to
exist. Since decreasing Gauss-Bonnet coupling corresponds to
decreasing string coupling $g_s$, the situation can tentatively be
interpreted as classical indication on the black hole -- string
transition. Previously the extremal dilaton black holes were studied
in the Einstein-frame version of the Gauss-Bonnet gravity. Here we
work in the string frame version of this theory with the S-duality
symmetric dilaton function as required by the heterotic string
theory.
\end{abstract}

\maketitle

\section{Introduction}The idea of correspondence between
black holes and massive string states suggested by Susskind
\cite{Suss} was elaborated in a number of papers \cite{horo,
Maldacena:1996ky, huri, damo, Corn, give, sen_global}. Though
naively the degeneracy of the string massive levels seems to
contradict the Bekenstein-Hawking area formula, it turns out that
there is an agreement between the geometric and the statistical
entropies for the critical mass at which the size of the string
configuration approaches the gravitational radius of the black hole
\cite{horo}. Some arguments in favor of the black hole -- string
transition were given in \cite{Corn} considering classical and
string pictures of the wave scattering. Here we would like to push
further  another argument based on the analysis  of the moduli space
of the curvature corrected extremal charged dilatonic black holes
\cite{chen}.

Extremal dilatonic black holes are especially suitable to probe the
conjectured string -- black hole transition  for the following
reason. To minimize uncertainties related to higher order quantum
corrections it is common to consider the BPS states usually
associated with black holes exhibiting a degenerate event horizon.
At the level of the Einstein-Maxwell-dilaton theory such black holes
are problematic since  their horizon radius tends to zero implying
vanishing Bekenstein-Hawking entropy. Meanwhile, from the string
theory side, the entropy is expected to be non-zero. Recently
understanding of this situation was improved by realizing that
curvature corrections \cite{Gros} to the Einstein action lead to
stretching the horizon  to a finite radius \cite{Sen:2005iz}, so
that the geometric entropy calculations fit to microscopic
predictions \cite{Moha,deWi} (both results being different from the
naive Bekenstein-Hawking value).

However, another problem was noticed in \cite{chen}: although the
entropies had been claimed to agree, the classical calculations were
based on the purely local considerations,  like Sen entropy function
approach \cite{Sen:2005iz}, which require solving  the Einstein
equations only in the vicinity of the event horizon. Meanwhile, to
be sure that one deals with the black hole indeed, global
integration of the Einstein equations is necessary which should
provide an extension of the local solutions to infinity. An attempt
of such a continuation in the Einstein-Maxwell-Gauss-Bonnet theory
with an arbitrary dilaton coupling \cite{chen} revealed existence of
the threshold behavior: global solutions cease  to exist for the
dilaton coupling above some critical value. This was interpreted as
an indication for the black hole -- string transition anticipated in
\cite{Corn}.

Here we would like to push this reasoning further and to mention its
possible relevance to the problem of the final stage of Hawking
evaporation. Note that the Gauss-Bonnet term is the most popular
candidate to imitate string corrections to the Einstein action
\cite{Pres}. The four-dimensional Newton constant $G_N$, the string
inverse tension $\alpha'$, and the string coupling $g_s$ are related
in four dimensions as $G_N\sim g_{s}^2\alpha'$. Thus, decreasing
string coupling $g_{s}$ implies decreasing  $G_N$. Since the
coefficient in front of  the Gauss-Bonnet term is dimensionless, the
decreasing $G_N$, entering the denominator of the Einstein term,
makes this term to dominate  the Gauss-Bonnet term. Technically,
however, it is more convenient to introduce a variable Gauss-Bonnet
coupling parameter $\alpha$ and to consider decreasing $\alpha$ for
fixed $G_N$. From the point of view of the black hole -- string
correspondence principle \cite{horo} the black hole case corresponds
to large $g_{s}^2$, hence, in our treatment, to large $\alpha$ (more
precisely, to large dilaton-renormalized $\alpha$ which we will call
later $\ta$). The existence of the lower bound for $\ta$ in the
black hole solution space then can be interpreted as a signal of the
black hole -- string transition.

Black hole solutions to the Einstein-Maxwell-dilaton theory with the
Gauss Bonnet term introduced in the Einstein frame were constructed
in \cite{chen}. They qualitatively confirm the above expectations.
However, from the string theory point of view, the Gauss-Bonnet term
has to be introduced in the string frame. Transformation of the
Einstein-frame version of the Gauss-Bonnet gravity to the string
frame leads to a lagrangian which differs from the string-frame
version of the Gauss-Bonnet gravity by the forth-order dilaton
derivative term. Therefore these two theories are not equivalent.
Here we investigate the extremal black holes in the second version,
adding the Gauss-Bonnet term in the string frame.

Another new feature of the present work is the use of the S-duality
symmetric dilaton function which is also motivated by string theory.
It is worth noting that, within the Einstein-Maxwell-dilaton model
without curvature corrections, electric and magnetic extremal black
holes look very different in the string frame. For magnetic black
holes the dilaton conformal factor cancels the divergent metric term
on the horizon, so in the string frame the solutions look regular.
For electric ones the singularity persists in the string frame too.
So the role of curvature corrections is expected to be more
significant for the electric solutions. Actual calculations show
that purely electric extremal black holes do not exist within the
string frame version of the Einstein-Maxwell-Gauss-Bonnet theory:
some magnetic charge is necessary. However one can distinguish the
electrically and the magnetically dominated dyons. Our
interpretation of the boundary in the parameter space as a signal of
the black hole -- string transition refers to the case of electric
dominance.
\section{String frame Gauss-Bonnet gravity}
We start with the following four-dimensional action:
\begin{equation}\label{eq:S2}
    \mathcal{S}=\frac{1}{16\pi G}\int S\Lef R+S^{-2}(\p
    S)^2-F^2\Rig \;d^4x\sqrt{-g}+\int \Delta\mathcal{L}\;d^4x\sqrt{-g},
\end{equation}
where
\begin{equation}\label{eq:GB1}
    \Delta \mathcal{L}=\frac{\alpha}{16\pi}\psi(S)\,L_{GB}, \quad
    L_{GB}=R^2-4R_{\mu\nu}R^{\mu\nu}+R_{\alpha\beta\mu\nu}R^{\alpha\beta\mu\nu}.
\end{equation}
The dilaton factor  $\psi(S)$ in front of the Gauss-Bonnet term
inheriting the S-duality symmetry of the heterotic string
\cite{sen_entropy, Cardoso:1999ur, harvey} can be presented as
\begin{equation}\label{eq:psi}
    \psi(S)=-\frac{3}{\pi}\ln(2S|\eta(iS)|^4),
\end{equation}
where the Dedekind $\eta-$function is defined on the complex plane
of $\tau$ as follows:
\begin{equation}\label{eq:psi2}
    \eta(\tau)\equiv e^{2\pi
    i\tau/24}\prod\limits_{n=1}^{\infty}(1-e^{2\pi i n\tau}).
\end{equation}
It is a modular form of the weight 1/2 satisfying the functional
equations
\begin{equation}
    \eta(-\tau^{-1})=\sqrt{-i\tau}\eta(\tau), \qd
    \eta(\tau+1)=e^{\frac{2\pi i}{24}}\eta(\tau),
\end{equation}
which represent the action of generators of the modular group. In
the general case $\tau=\mathrm{a}+iS$ where $\mathrm{a}$ is the
axion field. Here we consider only static spherically symmetric
configurations for which it  is consistent to set the axion
$\mathrm{a}=0$. In this truncated theory the function $\psi(S)$ is
invariant under the discrete $S$-duality   $S\to S^{-1}$. Note that
asymptotically for large $S$ it is a linear function of the dilaton:
$\psi(S)\to S$.

Consider the following parametrization of the static spherically
symmetric space-time:
\begin{equation}\label{eq:g1}
    ds^2=-w(r)\sigma(r)^2dt^2+\frac{dr^2}{w(r)}+\rho(r)^2d\Omega_{2}^2,
\end{equation}
and the ansatz for the gauge field:
\begin{equation}\label{eq:gauge}
    A=-f(r)dt-m\cos{\theta}d\vf,
\end{equation}
with $m$ denoting  magnetic charge.  Substituting this ansatz into
the action we obtain  the following one-dimensional lagrangian
\begin{eqnarray}\label{eq:L2}
    \mathcal{L} = S\left[ \sigma' w \rho\rho' +
    \frac{\sigma}{2}(\rho'(w\rho)'+1)\right]+
  S'\left(\sigma w \rho\rho'+\frac{\sigma
  w'\rho^2}{4}+\frac{\sigma'w\rho^2}{2}\right)&&\nn\\-
  \frac{\alpha\psi'}{\sigma}(w\sigma^2)'(w\rho'^2-1)+
  Sw\sigma\rho^2\frac{S'^2}{4S^2}+
  S\Lef\frac{\rho^2f'^2}{2\sigma}-\frac{\sigma
  m^2}{2\rho^2}\Rig,&&
\end{eqnarray}
where primes  denote   derivatives with respect to $r$. The
corresponding Euler-Lagrange equations  are rather complicated
except that for the vector field which can be integrated once to
give
\begin{equation}\label{eq:gauge32}
    f'=\frac{g}{\rho^2 S},
\end{equation}
where $g$ is an electric charge parameter defined on the horizon (an
asymptotical definition of the electric charge differs from this by
the dilaton exponent).  Therefore, after fixing the gauge
$\sigma=1$, the problem is reduced to three second-order ordinary
differential equations for three unknown functions $w, \rho, S$ and
a constraint equation.

One can investigate the residual symmetries of the lagrangian.
Consider first the dilaton shift (rescaling of $S$): the EMD part is
invariant under the following transformation with the global
parameter $\beta$:
\begin{equation}\label{eq:sym1}
    S\to \beta S\quad w\to\beta^4 w,\quad
    \rho\to\frac{\rho}{\beta},\quad
    r\to \beta r,\quad g\to g,\quad m\to \frac{m}{\beta}.
\end{equation}
The invariance of the GB term depends on the behavior of the
function $\psi(S)$. Another symmetry comes from the rescaling of the
electric  charge (with the global parameter $\gamma$) which causes
the following symmetry of the full lagrangian:
\begin{equation}\label{eq:sym2}
g\to \gamma g,\quad m\to \gamma m,\quad w\to\frac{w}{\gamma^2},
\quad \rho\to\gamma\rho, \quad\alpha\to\gamma^2\alpha.
\end{equation}
These symmetries will be used to clarify  nature of  parameters of
the  numerical solutions obtained below.
\section{Local solutions near the horizon}
To construct the extremal black hole solution (by extremality we
mean the $AdS_2\times S^2$ structure of the event horizon) we first
construct the local series solution in the vicinity of the event
horizon $r=r_0$. We use the gauge $\sigma=1$. The desired solution
must satisfy the conditions:
\begin{equation}\label{eq:series}
    w(r_0)=w'(r_0)=0,\qd \rho(r_0)=\rho_0>0, \qd \rho' (r_0)>0.
\end{equation}
 Thus the expansions must be on the form:
\begin{equation}
  w = \sum\limits_{n=2}^\infty w_n x^n, \qd
  \rho = \sum\limits_{n=0}^\infty \rho_n x^n, \qd
  S = \sum\limits_{n=0}^\infty p_n x^n, \qd
  \psi = \sum\limits_{n=0}^\infty t_n x^n,\label{eq:series4}
\end{equation}
where $x\equiv (r-r_0)$. Recall that  $\psi$ is  a function of the
dilaton field $S$. Therefore all $t_n$ are the functions of $p_i,\:
i=0..n$, explicitly one has
\begin{equation}\label{eq:series5}
    \psi=\Psi_0+\Psi_1 p_1 x+\Lef\frac{\Psi_2}{2}p_{1}^2+
    \Psi_1 p_2\Rig x^2+\Lef\frac{\Psi_3}{6}p_{1}^3+\Psi_2 p_1
    p_2+\Psi_1 p_3\Rig x^3+...,
\end{equation}
where  $\Psi_i=\Psi_i(p_0)=\psi^{(i)}(S)|_{S=p_0}$  are the $i-$th
derivatives of $\psi$ depending only on $p_0$. Substituting these
expansions into the equations of motion one obtains the sequence of
constraints on the coefficients. In the leading order we get the
relations between
  $g, \: m,\: \alpha$. Since we wish to determine the
range of $\alpha$ for which the black hole solutions exist, we
consider $\alpha$ as a variable parameter  on equal footing with the
electric and magnetic charges.  It is convenient to present  the
magnetic charge as $m=g\sqrt{u^2-1}/p_0$ with $u$ defined on the
semi-axis $u^2>1$, the remaining parameters will look simpler  in
terms of $u$. To avoid using an explicit dependence $\Psi_i(p_0)$
one can also leave
 $p_0$ as a  parameter and
express the expansion coefficients  as  functions of $(g,\:
u,\:p_0,\:\Psi_i(p_0))$, three of which   $g$, $u$ and $p_0$  are
independent.

In the leading order we find the following relations:
\begin{equation}\label{eq:series6}
    \alpha=\frac{g^2(2-u^2)}{4\Psi_1 p_{0}^2},\qd \rho_0=\frac{g
    u}{p_0},\qd w_2=\frac{p_{0}^2}{g^2u^2}.
\end{equation}
This corresponds to the Bertotti-Robinson metric $AdS_2\times S^2$
on the horizon with the curvature radius of the anti-de Sitter
sector coinciding with the radius of the two-sphere:
\begin{equation}\label{eq:ber}
    ds^{2}_{H}=-w_2x^2dt^2+\frac{dx^2}{w_2x^2}+\rho_{0}^2d\Omega_{2}^2,
\end{equation}
since $\rho_{0}^2=1/w_2$. With account for the second expansion
terms we have:
\begin{eqnarray}
  \rho &=& \frac{g u}{p_0}-\frac{gp_1(2\Psi_1+2\Psi_1u^2-\Psi_2p_0u^2)}{4\Psi_1 u p_{0}^2}x+O(x^2),\nn\\
  S &=& p_0+p_1 x+O(x^2), \nn\\
  w &=& \frac{p_{0}^2}{g^2u^2}x^2+\frac{\0
  p_1(2\Psi_1+2\Psi_1u^2-3\Psi_2p_0u^2)}{6g^2u^4\1}x^3+O(x^4).\label{eq:series7}
\end{eqnarray}
One can notice that the electric charge $g$ is just a scale factor
for the radial metric function and the gauge field: the
transformation (\ref{eq:sym2}) with $\gamma=1/g$ will remove this
parameter from the expansions. We can use this property to achieve a
correct asymptotic form of the solution at spatial infinity.
\section{Extending solutions to infinity}
We are interested in asymptotically flat metrics  such that
\begin{equation}\label{eq:series2}
    w(r)\to 1
    ,\quad \rho'(r)\to 1
    \qd {\rm as}\quad r\to \infty.
\end{equation}
We demand the metric to be asymptotically flat in the Einstein frame
as well, what implies  that the dilaton must be finite at infinity.
Consider a solution with the constant asymptotic value $w_\infty$ of
the metric component $-g_{00}=w $ at infinity. From the equations of
motion one has $w_\infty{\rho_{\infty}'}^2=1$, therefore an
asymptotic behavior of $\rho_{\infty}'$ is determined by $w_\infty$.
From the relations (\ref{eq:sym2}) it is clear that in order to pass
to the solution with another asymptotic value $\hat{w}_\infty$ one
should replace $g$ with $\hat{g}=g\sqrt{w_{\infty}/\hat{w}_\infty}$.
The dilatonic derivative coefficient $p_1=S'(r_0)$ enters the
expansions only in some combination with $x$, thus being the scale
factor as well. It can be removed by the transformation $x\to p_1
x,\: w\to p_{1}^2w$.

In the case of the linear function $\psi(S)=S$ one has $\1=1$ and
all other $\Psi_i$ vanish. Then the parameter $\0$ is a scale factor
too. Indeed, now the transformation (\ref{eq:sym1}) acts on the GB
term in the same way as on the EMD term if we rescale
$\alpha\to\alpha/\beta$, leading to the symmetry of the full
lagrangian under the dilaton shift. Note that the conditions
(\ref{eq:series2}) at spatial infinity are not invariant under the
dilaton shift, and the Minkowskian  asymptotic behavior can be
reached only for a unique value of $\0$ if the other parameters are
fixed. Also, for a purely electric configuration ($u=1$) all the
parameters will act as scale factors. This is the reason why we are
motivated to investigate more general dyonic systems.

 The solutions with $u=\sqrt{2}$ are possible only for  vanishing
$\alpha$. This is exactly the condition on the charges  $m=g/\0$ for
which the extremal limit of the EMD black hole  exists
\cite{Gibbons, Poletti:1995yq}. It has the $AdS_2\times S^2$ horizon
and the Minkowski asymptotic behavior. In the \emph{string} frame
the solution takes the form
\begin{eqnarray}
  \rho &=& \gamma^{-1}\left[\mathcal{W}_0\left(\delta e^{\gamma x+\delta}\right)+1\right],
  \quad \gamma=\frac{\8\0}{\sqrt{2}g(\8-\0)},\quad \delta=\frac{\0}{\8-\0}, \nn \\
  S &=& \8\rho',\qd
  w\rho'^2=\left(1-\frac{\sqrt{2}g}{\0\rho}\right)^2.
\end{eqnarray}
Here $\8$ is the asymptotic value of the dilaton at spatial
infinity, and $\mathcal{W}_0$ denotes the real branch of the Lambert
$\mathcal{W}$-function which satisfies the following functional and
differential equations:
\begin{equation}
    z=\mathcal{W}(z)e^{\mathcal{W}(z)},\qd
    z\frac{d\mathcal{W}}{dz}=\frac{\mathcal{W}}{\mathcal{W}+1}.
\end{equation}

Let us discuss the S-duality symmetry of the back hole solutions we
are looking for. The action of the S-duality transformations on the
charges and the dilatonic background on the horizon in the general
case of our model with an axion present is \cite{sen_sdual,
Sen:2005iz}:
\begin{equation}\label{a:sdual}
    \left(%
\begin{array}{c}
  g' \\
  m' \\
\end{array}%
\right)=\left(%
\begin{array}{cc}
  l & n \\
  r & s \\
\end{array}%
\right)\left(%
\begin{array}{c}
  g \\
  m \\
\end{array}%
\right),\qd
\mathrm{a}_{0}'+i\0'=\frac{l(\mathrm{a}_{0}+i\0)+n}{r(\mathrm{a}_{0}+i\0)+s};\qd
(l,\: n,\: r, \: s)\in \mathbb{Z},\quad ls-nr=1.
\end{equation}
If we set zero the horizon value of the axion
$\mathrm{a}_{0}'=\mathrm{a}_0=0$, we will get the constraints on the
parameters of the transformation $(l,\: n,\: r, \: s)$. There are
two possibilities:
\begin{eqnarray}
  &&n=r=0,\quad s=1/l \qd\Rightarrow\qd g'=lg,\quad m'=m/l,\quad \0'=l^2\0; \label{a:sdual2}\\
  &&l=s=0,\quad r=-1/n \quad\;\Rightarrow\qd g'=nm,\quad m'=-g/n,\quad
  \0'=n^2/\0. \label{a:sdual3}
\end{eqnarray}
The first transformation is a kind of rescaling
(\ref{eq:sym1},\ref{eq:sym2}) with $\beta=l^2,\:\gamma=l$. The
second transformation is the discrete S-duality: it interchanges
the electric and magnetic charges and  inverts  the dilaton.

Now consider the unique value of the parameter $u=\sqrt{2}$, for
which the horizon  expansions  are compatible with switching off the
Gauss-Bonnet term. The S-duality transformation acts on $u$ as
follows:
\begin{eqnarray}
  (\ref{a:sdual2}) &\Rightarrow& u'=\sqrt{\frac{m'^{2}\0^{2}}{g'^{2}}+1}=\sqrt{\frac{m^{2}\0^{2}}{g^{2}}+1}=u, \nn\\
  (\ref{a:sdual3}) &\Rightarrow&
  u'=\sqrt{\frac{m'^{2}\0'^{2}}{g'^{2}}+1}=\sqrt{\frac{g^{2}}{m^{2}\0^2}+1}.
\end{eqnarray}
If $u=\sqrt{2}$, then $m^{2}\0^{2}/g^{2}=1=g^{2}/(m^{2}\0^2)$ and
$u'=u$. Therefore the surface $u(g,m,\0)=\sqrt{2}$ in the parameter
space maps into itself under  the S-duality.

Our goal is to find the region of parameters for which
asymptotically flat extremal black holes do exist. From numerical
calculations it follows that this is only possible in some region of
the parameters $m, \alpha$ bounded from below.  Note that the
transformation to the Einstein frame reads:
\begin{equation}\label{eq:Einst1}
    ds_{E}^2=Sds^2=-wSdt^2+\frac{d\tilde{r}^2}{wS}+\rho^2Sd\Omega_{2}^2,\qd
    \mbox{where}\quad d\tilde{r}=Sdr.
\end{equation}
So the solution corresponding to the parameter
$\hat{g}=g\sqrt{w_\infty S_\infty}$ asymptotically will satisfy the
relations $w_\infty=1/S_\infty,\: \rho_{\infty}'=\sqrt{S}$ which
correspond to the Minkowski space in the Einstein frame. It turns
out that the local solutions defined by the series expansions on the
horizon can be extended to spatial infinity not for all values of
the parameters. As we have discussed, the family of solutions we are
interested in is characterized by two free parameters, if one
considers the Gauss-Bonnet coupling $\alpha$ as a parameter. Since
in our treatment the dilaton $S$ generically is not equal to one at
infinity, it is more appropriate to use a dilaton-renormalized GB
coupling constant $\ta=\alpha \Psi_0 $ instead of $\alpha$. As the
second parameter we use the quantity $u$ replacing the magnetic
charge.

\begin{figure*}[p]
\hbox to\linewidth{\hss%
 \psfrag{a}{\huge{$u$}}
 \psfrag{z}{\huge{$\ta^{1/4}$}}
    \resizebox{10cm}{6.5cm}{\includegraphics{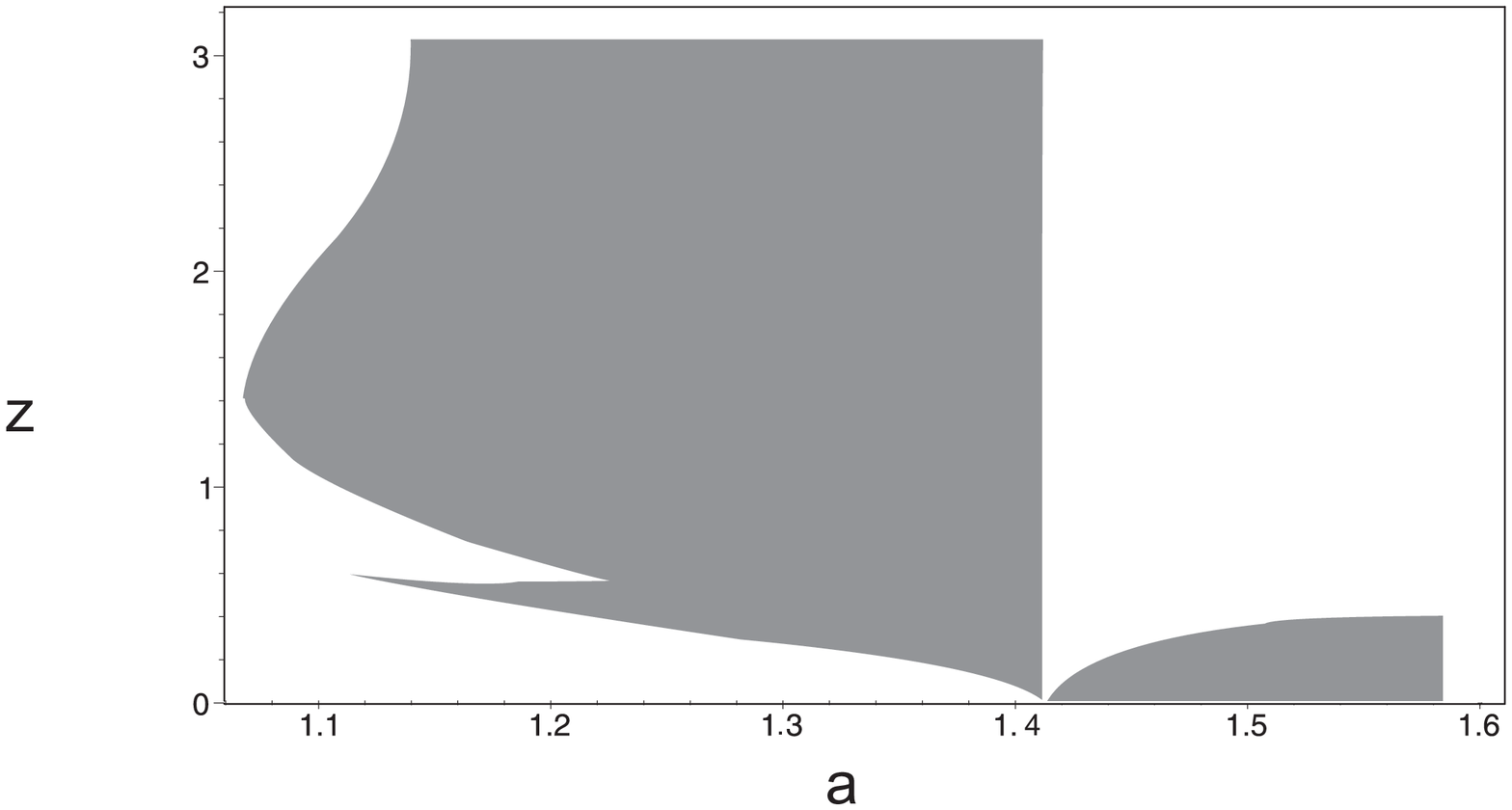}}
\hss} \caption{\small The domain of existence of regular black hole
solutions in the plane of the magnetic parameter $u$  and the GB
coupling~$\ta$.} \label{fig:planef}
\end{figure*}

We explored the regions in the parameter plane of $\ta,\; u$  for
which regular solutions exist both for the linear and for the
self-dual forms of the function $\psi(S)$. Numerical calculations
reveal that for the linear dilaton function $\psi(S)=S$ there are no
regular black hole solutions at all. But for the S-duality symmetric
function $\psi(S)$   given by (\ref{eq:psi}) such solutions {\em do
exist} in the region $\ta\geq\ta_{\rm min}$ for $u<\sqrt{2}$, and in
the region $\ta\leq\ta_{\rm max}$ for $u>\sqrt{2}$, as shown  in
Fig.~\ref{fig:planef}. Passing from $u$ to the electric and magnetic
charges one finds that the first region corresponds to the
electrically dominated  configurations  for which $m\0<g$, while the
second --- to the magnetically dominated ones. Therefore the
electrically dominated black holes with $AdS_2\times S^2$ horizons
exist when the (dilaton renormalized) GB coupling $\ta$ exceeds some
minimal value, while the magnetically dominated black holes exist
for $\ta$ bounded from above. In what follows we will concentrate on
the first case as potentially relevant to our discussion.

The ADM mass and the dilaton charge  can be extracted from the
 expansions at spatial infinity in the Einstein frame:
\begin{eqnarray}
  w_E &=& 1-\frac{2M}{\hat{r}}+O(\hat{r}^{-2}),\nn \\
  S &=& S_\infty+\frac{2S_\infty D}{\hat{r}}+O(\hat{r}^{-2}).
\end{eqnarray}
It turns out  that in the limit $\ta\to\infty$ the mass $M$ remains
finite while the dilaton charge $D$ and the asymptotic value of the
dilaton  $S_\infty$ diverge. Two other physical parameters
appropriate to global solutions are the  electric and magnetic
charges $Q$ and $P$ defined asymptotically by integration of the
corresponding fluxes  over the large sphere:
\begin{equation}\label{eq:charges}
    Q=\frac{g}{\sqrt{S_\infty}},\qd
    P=m\sqrt{S_\infty}.\qd
\end{equation}
They differ from the charges  defined on the horizon by the
asymptotic value of the dilaton. In the limit $\ta\to\infty$ the
electric charge $Q$ tends to a constant while the magnetic charge
$P$ diverges.

\begin{figure*}[p]
\hbox to\linewidth{\hss%
 \psfrag{z}{\huge{$\ta^{1/4}$}}
 \psfrag{b}{\LARGE{$C_{BPS}$}}
 \psfrag{u}{\huge{$u=1.25$}}
 \psfrag{v}{\huge{$u=1.39$}}
    \resizebox{10cm}{6.5cm}{\includegraphics{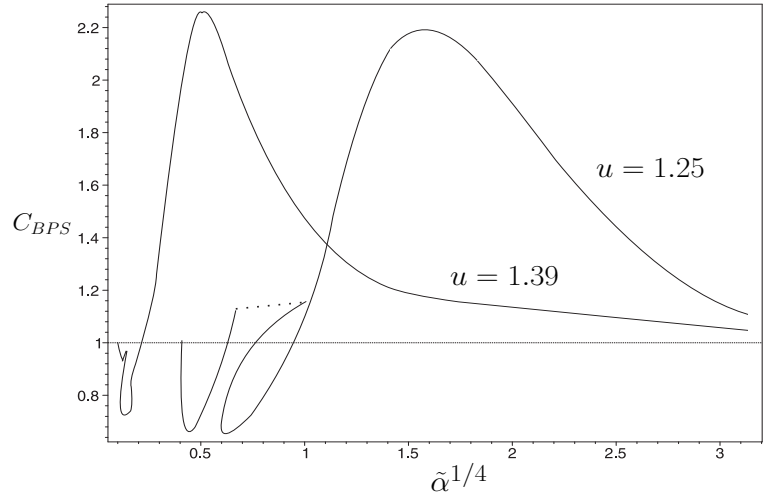}}
\hss} \caption{\small The ratio $C_{BPS}\equiv (Q^2+P^2)/(M^2+D^2)$
as a function of the GB coupling $\ta$ for $u=1.25$ and $u=1.39$.
Dotted line shows
the interval where regular solutions do not
exist.} \label{fig:bps}
\end{figure*}

Near the boundaries  of the allowed region  of  $\ta$, namely for
$\ta\to\ta_{\rm min}$ and $\ta\to\infty$, one observes a peculiar
relation between the asymptotic charges and the ADM mass. Recall,
that in the EMD theory without curvature corrections the BPS limit
corresponds to the following condition  \cite{Gibbons, bps}:
\begin{equation}\label{eq:bps1}
    M^2+D^2=Q^2+P^2.
\end{equation}
When $\ta\to\infty$ (large curvature corrections), the dilaton
charge $D$ and the magnetic charge $P$ diverge while the mass $M$
and the electric charge remain $Q$ constant. In the case of  small
curvature corrections,  $|D|$ and $P$ are small with respect to $M$
and $Q$. To describe the results of numerical calculations we
introduce the ratio
\begin{equation}\label{eq:bps2}
    C_{BPS}=\frac{Q^2+P^2}{M^2+D^2}
\end{equation}
and explore its variation with  growing  $\ta$ from $\ta_{\rm min}$
to infinity. The results for two fixed values of the magnetic
parameter $u=1.25$ and $u=1.39$ are shown in Fig.~\ref{fig:bps}. It
turns out that in  both  limits $\ta=\ta_{\rm min}$ and
$\ta\to\infty$ the BPS condition (\ref{eq:bps1}) of the supergravity
without curvature corrections is fulfilled. Thus the family of
curvature corrected black hole solutions is bounded by two
quasi-BPS states. The lower limit is the charged black hole with the
mass $M$ and the absolute value of the electric charge $|Q|$ of the
same order  and significantly larger than the dilaton charge $|D|$
and the magnetic charge $|P|$. The upper limit is the black hole
with extremely strong magnetic and dilaton fields. With $u$
approaching $\sqrt{2}$, the plot shrinks to the abscissa. The limit
$u=\sqrt{2}$ corresponds to the single BPS solution when the range
of $\ta$ shrinks to the point $\ta=0$.

The entropy of our black holes can be calculated using the Sen
entropy function approach \cite{Sen:2005iz, sen_entropy}. The result
is:
\begin{equation}\label{eq:fen11}
    \mathrm{S}=\pi\rho_{E}^2+4\pi\ta.
\end{equation}
The first term is exactly the Bekenstein-Hawking entropy, while the
second term describes  curvature corrections. Evaluating $\ta$ in
terms of the expansion parameters we obtain
\begin{equation}\label{eq:fen12}
    \mathrm{S}=\pi\rho_{E}^2\left[1+\frac{\Psi_0}{u^2\0\1}
    \left(2-u^2\right)\right].
\end{equation}
\section{Conclusions}
Let us briefly summarize the results obtained. Our aim was to
investigate the domain of an effective curvature coupling parameter
$\ta$ for which the system admits the extremal black hole solutions
with the horizon geometry $AdS_2\times S^2$. We have found that if
$\ta$ exceeds some minimal value $\ta_{\rm min}$, the black holes
exist which are endowed with both electric and magnetic charges. For
lower values of $\ta$ in the allowed region the electric charge is
dominant, while for large enough $\ta$ the solutions are
magnetically-dominated. On the boundary $\ta\ra \ta_{\rm min}$ and
in the asymptotic region  $\ta \ra \infty$ the mass and the charges
satisfy the same BPS condition, as singular extremal black holes in
the theory without curvature corrections. This feature is similar to
that observed previously in the Einstein-frame Gauss-Bonnet
gravity~\cite{chen}.

We think that the existence of the lower $\ta$-boundary for
solutions in the region $u<\sqrt{2}$ can be traced to the
conjectured black hole -- string transition. Indeed, the string
coupling parameter $g_s$ is large for large $\ta$ in which case the
system admits the black hole solutions. With decreasing $g_s$ one
observes disappearance of the black hole solutions while the mass
and the charges of the configuration are still finite. This is what
is expected for the transition region. One can also argue that in
the process of evaporation the mass of the black hole decreases and
at some moment the parameters of the solution leave the allowed
domain, so one enters into the transition region. However this
hypothesis must be investigated in more detail taking into account
an actual evolution of  charges during the evaporation process.

 \begin{acknowledgments}
The authors are grateful to Chiang-Mei-Chen and Dmitri Orlov for
helpful discussions. The paper was supported by the RFBR grant
08-02-01398-a.
 \end{acknowledgments}


\end{document}